# Broadband Terahertz Generation from Metamaterials Pumped at Telecomm Wavelengths

Liang Luo[1], Ioannis Chatzakis[1†], Jigang Wang[1*], Fabian B. P. Niesler[2], Martin Wegener[2], Thomas Koschny[1], and Costas M. Soukoulis[1, 3*]

[1] Department of Physics and Astronomy and Ames Laboratory—U.S. DOE, Iowa State University, Ames, Iowa 50011, USA

[2] Institute of Applied Physics, Institute of Nanotechnology, and DFG-Center for Functional Nanostructures (CFN), Karlsruhe Institute of Technology (KIT), 76128 Karlsruhe, Germany

[3] Institute of Electronic Structure and Lasers (IESL), FORTH, 71110 Heraklion, Crete, Greece

† Present address: Department of Materials Science and Engineering, Stanford University, Stanford, California 94305, USA and Stanford institute for Materials and Energy Science, SLAC National Accelerator Laboratory, Menlo Park, California 94025, USA

**The Terahertz spectral regime ranging from about 0.1 to 15 THz (1 THz = $10^{12}$ Hz), is one of the least explored yet most technologically transformative spectral regions[1-5]. One key current challenge is to develop efficient and compact THz emitters/detectors with a broadband and gapless spectrum that can be tailored for various pump photon energies. Recently, the development of metamaterials composed of split ring resonators (SRRs) has enabled to tailor resonant optical nonlinearities from the THz to the infrared and visible regions[6,7]. While nonlinear metamaterials have been actively pursued[8-16], THz generation from any type of metamaterial has not been reported. Here we demonstrate efficient single-cycle, broadband THz generation, ranging from about 0.1 to 4 THz, from a thin layer of SRRs with few tens of nanometers thickness by pumping at**

**the telecomm-wavelength of 1.5 μm (200 THz). The THz emission arises from exciting the magnetic-dipole resonance of SRRs and quickly decreases under off-resonance pumping. This, together with pump polarization dependence and power scaling of the THz emission, identifies the role of optically induced nonlinear currents in SRRs. We reveal a giant sheet nonlinear susceptibility ~$10^{-16}$ m$^2$/V that far exceeds thin films and bulk non-centrosymmetric materials such as ZnTe[17-19].**

The challenge to develop and apply THz light sources merges different disciplines of fundamental science and technology, ranging from ultrafast nonlinear optics, condensed matter and materials physics to optoelectronics and microwave photonics. The emerging THz technologies, such as quantum-cascade lasers[1-3], ultrafast photoconductive switches[4], have enabled various THz spectroscopy/imaging/sensing developments and have offered perspectives, amongst others, for pushing the GHz switching speed limit of today's logic/memory/wireless communication devices into the THz regime[4,5]. In regard to broadband THz sources, a major recent progress[20] is based on nonlinear optical rectification in inorganic crystals such as ZnTe/GaP/GaAs/GaSe/DAST/LiNbO3, pumped by femtosecond laser pulses combined with field-resolved detection via electro-optic sampling using similar crystals. However, issues in these crystals are: (1) strongly absorbing longitudinal optical phonon bands that lead to a gap in the THz spectrum in the Reststrahlen region; (2) subtle quasi-phase-matching conditions that require locking phase velocity of the THz emission to group velocity of the optical pump, which restricts the accessible pump photon energies to a fixed and narrow range in order to increase the coherent length.

Recently, there have been efforts on thin THz emitter/detector crystals of 10-30 μm thickness[20,21] to increase the THz emission bandwidth by leveraging above-mentioned restrictions and to enable integration with current micro-/opto-electronics technology. However, the relatively small nonlinear susceptibilities of inorganic emitters limit the THz emission intensity. In this regard, investigating single nanometer thickness metamaterials exhibiting artificial *optical* magnetism –i.e., sustaining circulating ring currents at optical frequencies – can meet the urgent demand for new nonlinear materials for optical rectification free from both quasi-phase-matching limitation and spurious THz absorption. This is mainly due to the coexistence of resonant nonlinearity[22-24] from magnetic dipoles and local electric-field enhancement in the narrow gaps of the SRRs, which together allows efficient and broadband THz radiation from emitters of strongly reduced thickness. In addition, tailoring the magnetic resonances of the metamaterial emitters allows for matching to essentially any desired pump photon energy. For example, this allows for integrating THz optoelectronics with high-speed telecommunications as the 1.3-1.5 μm range is not ideal for the above inorganic crystals. Some other nonlinear media have also been explored although they mostly suffer from either poor stability or limited bandwidth, e.g., bulky ambient air-plasma generation normally with large shot-by-shot fluctuations[25], electrically-biased InGaAs photoconductive switches that not only require extra voltage driving source but also have limited bandwidth of ~1 THz[26]. The same narrow THz emission bandwidth has also been seen from the fs-laser-accelerated photoelectrons in nano-plasmonic structures[27], with intensity still ~4 orders smaller than optimal ZnTe crystals, which limits the perspective for their applications to THz emitters[28].

In this Letter we demonstrate efficient THz emission up to 4 THz from optical rectification in a single SRR layer of 40 nm in thickness, based on resonant photoexcitation of the magnetic-dipole resonance centered around 1500 nm wavelength using 140 fs laser pulse. The strong THz emission intensity from the metamaterial is on the same order as optimal ZnTe crystals that are thousand times thicker, revealing a gigantic resonant second-order sheet nonlinear susceptibility of SRRs ~$10^{-16}$ $m^2$/V, which is three orders of magnitude higher than the typical surface and sheet values of bulk crystals and thin films[17-19].

Fig. 1a schematically illustrates the experiment and shows key elements. A Ti:sapphire amplifier is used to drive the whole setup with center wavelength 800 nm, pulse duration 35 fs, and repetition rate 1 kHz. The main portion of the output from the amplifier is used to pump an optical parametric amplifier (OPA) to produce tunable near-infrared (NIR) radiation from 1100-2600 nm of about 140 fs pulse duration. The NIR radiation from the OPA is used as a generation beam to pump the metamaterial emitter made of a single layer of 40 nm thin SRRs. The generated THz pulses are focused onto a ZnTe or GaSe crystal by a parabolic mirror for electro-optic sampling by a small portion of the amplifier output. A wire grid polarizer is used to measure the polarization of the generated THz pulses (supplementary). An electron micrograph of the metamaterial SRR array is shown in the magnified frame at the bottom left of Fig. 1a. The sample has been fabricated using electron-beam lithography and high-vacuum evaporation of gold, followed by a lift-off procedure. The single layer of SRRs is 40 nm thick with a square lattice constant of 382 nm. The big yellow SRR on the left illustrates the geometrical parameters used for the numerical calculations. Three small yellow SRRs are overlaid in the middle of SRR array micrograph in order to compare the designed SRR geometry with the actual SRR dimension.

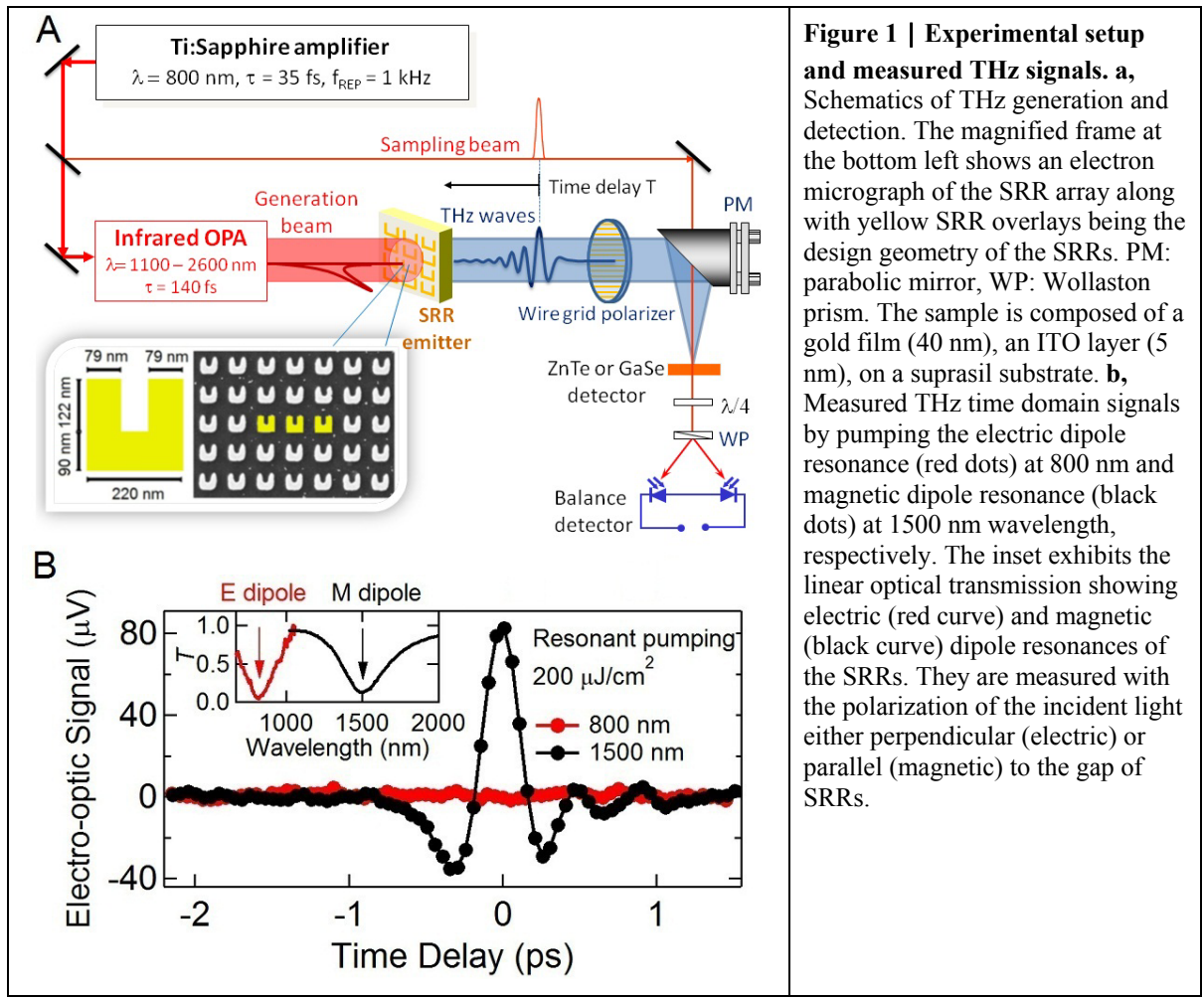

**Figure 1 | Experimental setup and measured THz signals. a,** Schematics of THz generation and detection. The magnified frame at the bottom left shows an electron micrograph of the SRR array along with yellow SRR overlays being the design geometry of the SRRs. PM: parabolic mirror, WP: Wollaston prism. The sample is composed of a gold film (40 nm), an ITO layer (5 nm), on a suprasil substrate. **b,** Measured THz time domain signals by pumping the electric dipole resonance (red dots) at 800 nm and magnetic dipole resonance (black dots) at 1500 nm wavelength, respectively. The inset exhibits the linear optical transmission showing electric (red curve) and magnetic (black curve) dipole resonances of the SRRs. They are measured with the polarization of the incident light either perpendicular (electric) or parallel (magnetic) to the gap of SRRs.

Typical time-domain THz traces, $E_{THz}$, during the 4 ps interval and for a pump fluence of 200 μJ/cm$^2$, are shown in Fig. 1b for two pump wavelengths: this clearly demonstrates THz generation *exclusively* from pumping 1500 nm (inset, black arrow) at the magnetic-dipole resonance (black dots), while there is negligible THz signal from pumping 800 nm (inset, red arrow) at the electric-dipole resonance of the SRRs (red dots). Note that the two resonances are excited with the linear polarization of the incident pump light either perpendicular (electric) or parallel (magnetic) to the gap of SRRs. In addition, we characterize the polarization state of the emitted THz pulses to be perpendicular to the gap of the SRRs, consistent with the second-order nonlinearity (see supplementary).

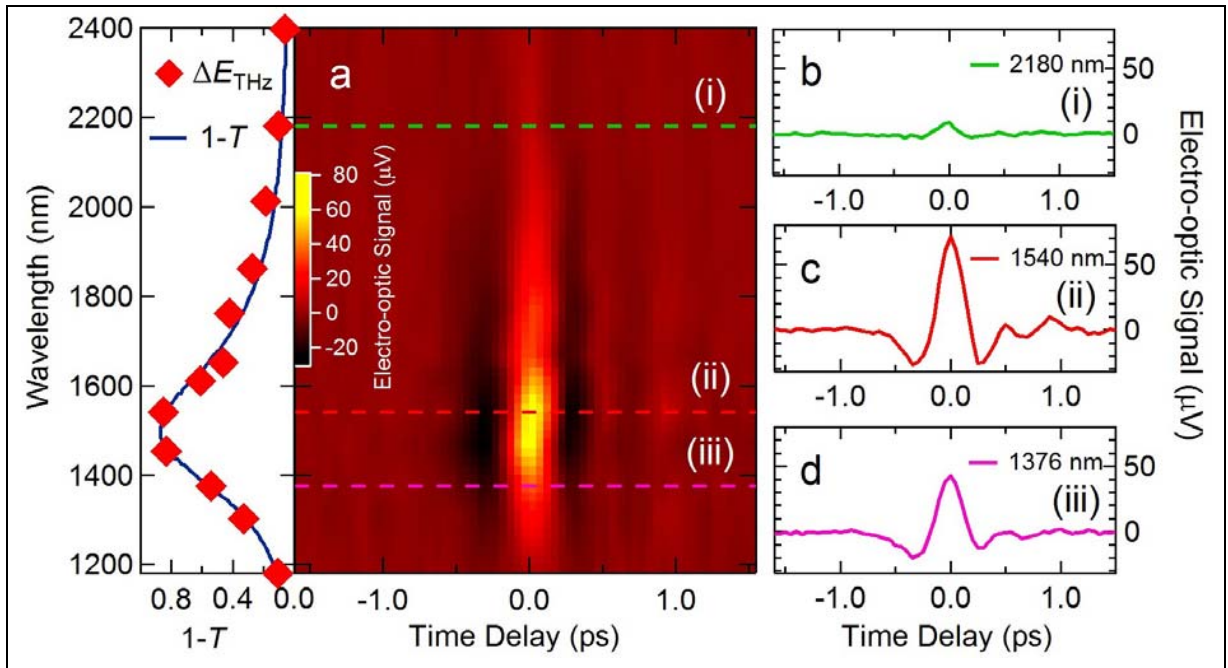

**Figure 2 | Pump-wavelength dependence of THz generation from the SRR metamaterial. a,** A two–dimensional false-color plot of the THz temporal traces for sweeping the pump wavelength across the magnetic-dipole resonance. The left-hand-side panel summarizes peak-to-peak amplitudes of THz emission (red diamonds) as a function of pump wavelength, together with the 1 - *T* curve (blue solid line). *T* is the measured transmission around the SRR's magnetic-dipole resonance. The three colored curves (green, red, magenta) indicate the cut positions of three time domain traces shown in **b-d** corresponding to 2180 nm, 1540 nm, and 1376 nm respectively. If the pump wavelength is detuned from the resonance, the signal amplitude quickly decreases and the line shape matches very well with the 1 - *T* curve.

To further investigate the nature of the THz emission, the false-color plot and the time-domain THz traces in Fig. 2 show the detailed excitation-wavelength dependence. Resonant photoexcitation of the magnetic-dipole resonance ~1540 nm leads to significant enhancement of the THz emission, as shown in Fig. 2a. The corresponding time-domain trace for the on-resonance pumping is plotted in Fig. 2c. The conversion efficiency quickly decreases under off-resonance pumping, which is shown by the reduction of the THz emission for excitation both below and above the magnetic-dipole resonance, at 2180 nm (Fig. 2b) and 1376 nm wavelengths (Fig. 2d), respectively. The peak-to-peak amplitudes of THz emission $\Delta E_{THz}$ (red diamonds) closely follow the absorption around the magnetic-dipole resonance, as seen by comparison with the 1-$T$ curve (blue, left side panel, Fig. 2a), where $T$ is the linear optical transmission of the sample. All of these observations corroborate the resonant THz generation from nonlinear currents induced by the electrically-coupled fundamental magnetic SRR resonance.

Figure 3a compares the peak-to-peak amplitude of THz electric field $\Delta E_{THz}$ versus pump power for the one-layer metamaterial emitter with 40 nm thickness (blue squares) with two ZnTe emitters with thicknesses of 1 mm (red diamonds) and 0.2 mm (magenta dots), respectively. The THz signals for the three emitters are measured at the same pump fluence and using the same ZnTe detector crystal of 1 mm in thickness. For the ZnTe emitters pumped at 1500 nm, coherence length $L_c$ for the THz emission is ~0.2 mm[29]. Increasing the crystal thickness beyond this value actually decreases the emission intensity, as seen for the 1 mm case. This effect is expected from the quasi-phase-matching in the ZnTe emitter crystal: after propagating $L_c$, the superposition of the emitted THz pulses from different sheets will have destructive interference in the ZnTe. Most intriguingly, Fig. 3a reveals a remarkably efficient THz generation from the single layer of SRRs with emission intensity on the same order as the optimal ZnTe emitters of 5000 times thicker: the SRRs generate THz radiation roughly 1/5 of the maximum from the 0.2 mm ZnTe. The emitted THz *electric field* scales linearly with the pump power which corroborates the optical rectification process. Our phase-resolved THz measurements from the SRR and ZnTe emitters allow us to unambiguously determine the second order nonlinear susceptibility of a single SRR layer to be $0.8 \times 10^{-16}$ m$^2$/V (supplementary), which has not been possible from prior nonlinear measurements such as second-harmonic generation (SHG)[22,23].

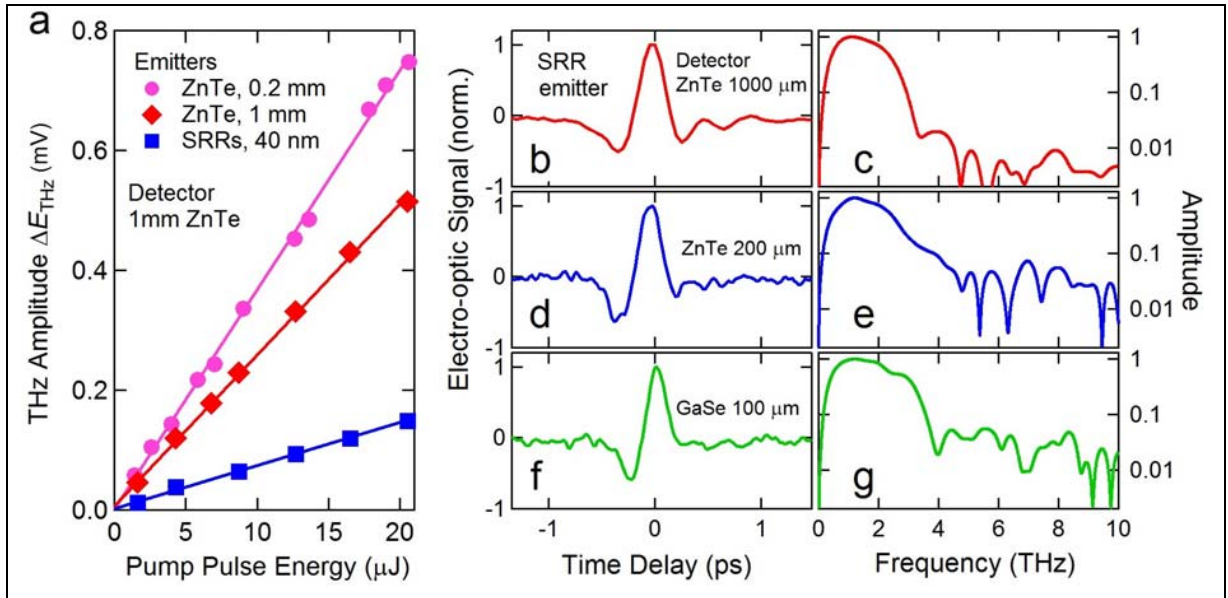

**Figure 3 | Comparison of emitters and measured spectral bandwidths. a,** The peak-to-peak amplitude of THz signal versus pump power measured with three different THz emitters: 0.2 mm ZnTe (magenta dots), 1 mm ZnTe (red diamonds) and SRR (blue squares) and the same detector of 1 mm ZnTe. The straight lines are from linear fitting. **b-g**, The normalized time-domain THz pulses and corresponding spectral amplitudes generated from the SRR emitter and three different detectors. **b**,**c**: 1 mm ZnTe detector (red); **d**,**e**: 0.2 mm ZnTe (blue); **f**,**g**: 0.1 mm GaSe (green).

Figures 3b-3g plot normalized temporal profiles of THz electric field pulses and corresponding spectral amplitudes from a single layer SRR emitter. The THz detector crystals are 1 mm ZnTe (red, Figs. 3b-3c), 0.2 mm ZnTe (blue, Figs. 3d-3e), and 0.1 mm GaSe detector (green, Figs. 3f-3g). This demonstrates the THz bandwidth of the SRR emitter up to 4 THz which is limited mostly by the excitation pulse duration ~ 140 fs (spectral width ~13 meV or 3.2 THz) and by the Reststrahlen region of the inorganic nonlinear detector crystals (centered ~5-6 THz) (supplementary). Our approach can potentially even generate much higher THz bandwidth by shortening the pump pulses because the single layer SRR emitter does not suffer from the intrinsic limitation of the Reststrahlen region in almost all inorganic THz emitters/sensors.

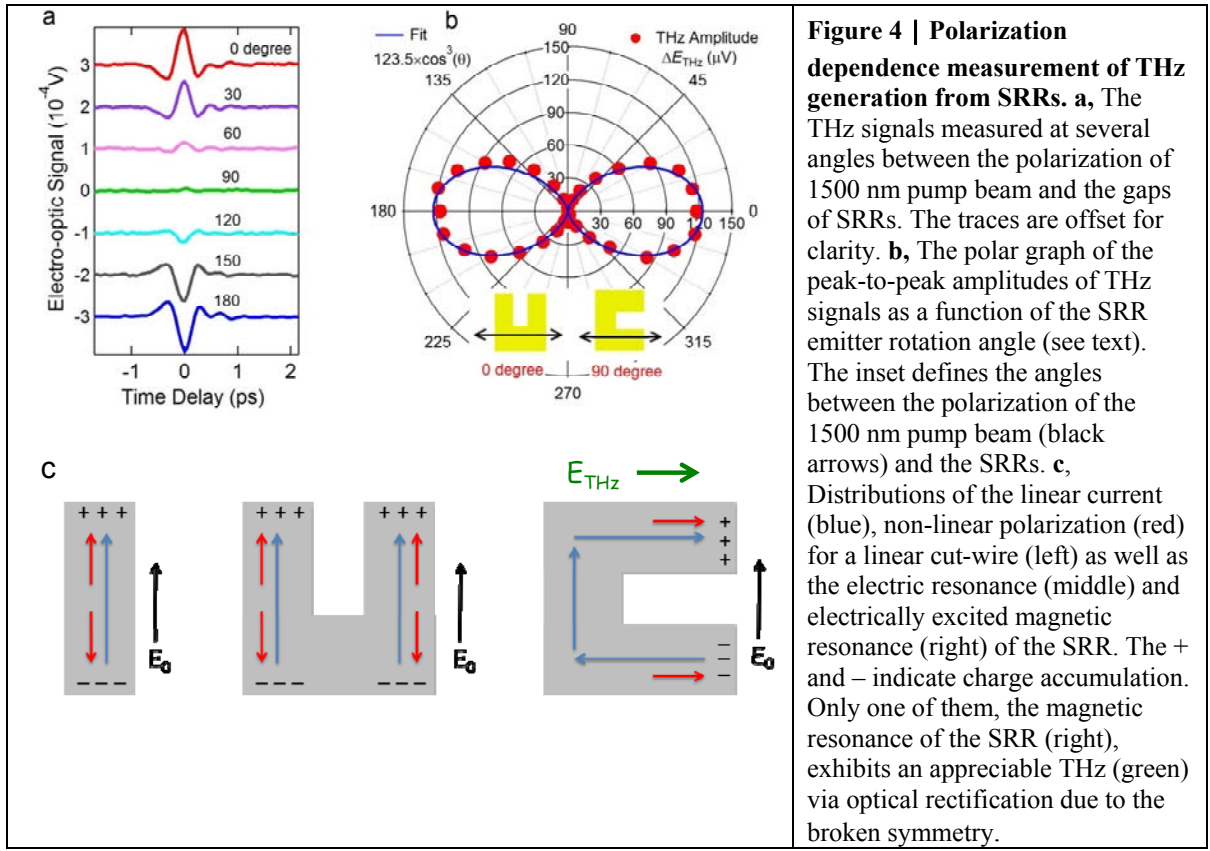

**Figure 4 | Polarization dependence measurement of THz generation from SRRs. a,** The THz signals measured at several angles between the polarization of 1500 nm pump beam and the gaps of SRRs. The traces are offset for clarity. **b,** The polar graph of the peak-to-peak amplitudes of THz signals as a function of the SRR emitter rotation angle (see text). The inset defines the angles between the polarization of the 1500 nm pump beam (black arrows) and the SRRs. **c,** Distributions of the linear current (blue), non-linear polarization (red) for a linear cut-wire (left) as well as the electric resonance (middle) and electrically excited magnetic resonance (right) of the SRR. The + and – indicate charge accumulation. Only one of them, the magnetic resonance of the SRR (right), exhibits an appreciable THz (green) via optical rectification due to the broken symmetry.

Fig. 4a shows the control of the amplitude and phase of THz emission by varying relative polarization by rotating the SRR emitter. As shown in the inset of Fig. 4b, we define the polarization angle to be 0 degree when the polarization of pump beam is parallel to the gap of the SRR, and 90 degrees for the orthogonal polarization. Fig. 4a plots 7 traces by varying the SRR emitters from 0 to 180 degree under 1500 nm pumping (traces are offset for clarity). It clearly shows that the THz emission amplitude oscillates starting from a maximum at 0 degree, via a zero crossing at 90 degrees, back to a maximum at 180 degrees with a π phase shift from the origin. The complete polarization dependence of the emission amplitude is shown as a polar graph in Fig. 4b where the peak-to-peak amplitude of THz signal is plotted as a function of the rotation angle. The amplitude is best fitted with a $\cos^3(\theta)$ function, where a $\cos^2(\theta)$ dependence is from the polarization between the pump beam and the gap of the SRR, and the other $\cos(\theta)$ originates from the detection ZnTe crystal. In short, the detector is aligned such that it measures vertically (horizontally) polarized THz pulses at maximum (minimum) efficiency and the efficiency follows a cos(θ) function[30].

Optical rectification, responsible for the observed THz emission, is induced by the broken symmetry in the second order nonlinear current distribution, which arises exclusively from the photoexcited magnetic-dipole resonance of the SRR. The second-order nonlinearities for the metallic response can be described by a hydrodynamic model[24] know as Maxwell-Vlassov theory. It treats the main non-linear second-order contributions originating from the $\mathbf{E}\,\mathrm{div}\,\mathbf{E}$ and $(\mathbf{j}\cdot\mathrm{grad})\mathbf{j}$ terms, which do not contribute in the bulk but do contribute on surfaces (supplementary). Both terms can induce the nonlinear current either parallel or antiparallel to the linear current induced by the external excitation in regions of increasing or decreasing surface charge, respectively, as illustrated in Fig. 4c. In the case of a simple straight nanorod (left panel), the radiations caused by the non-linear current contribution (red) in the two regions of surface charge accumulation (i.e., both ends of the nanorod) are out of phase and thereby interfere destructively. Therefore, no emission is observable in the far field. Similarly, the symmetry of the current distribution at the electric dipole resonance of the SRR (middle panel) cannot induce radiation either. If, however, one bends the nanorod into an U-shaped SRR and excites the magnetic resonant mode (right panel) that has a continuous current (without nodes) around the SRR ring, the non-linear current in both arm are now parallel and their radiated fields interfere constructively. This leads to the observed THz emission in the far field perpendicular to the SRR gap (green line). It should be noted that, although the mechanism also holds for the SHG reported previously[22,23], the current phase-resolved THz results allow us to solve the completely different pressing challenges in THz opto-electronics and to provide novel insights in nonlinear optics of metamaterials, e.g., quantitatively reveal the nonlinear susceptibility of the SRRs previously inaccessible and phase reversal of THz emission.

**Acknowledgments**

Work at Ames Laboratory was partially supported by the U.S. Department of Energy, Office of Basic Energy Science, Division of Materials Sciences and Engineering (Ames Laboratory is operated for the U.S. Department of Energy by Iowa State University under Contract No. DE-AC02-07CH11358) (experiments) and by the U.S. Office of Naval Research, Award No. N00014-10-1-0925 (theory). J.W. also acknowledge support by the National Science Foundation (contract no. DMR-1055352). The Karlsruhe team acknowledges support by the DFG, the State of Baden-Württemberg, and the Karlsruhe Institute of Technology (KIT) through the DFG-Center for Functional Nanostructures (CFN) within subproject A1.5.

**Correspondence**

To whom correspondence should be addressed. E-mail: C. M. S. (soukoulis@ameslab.gov); J. W. (jgwang@iastate.edu)

Supplementary information for

# Broadband Terahertz Generation from Metamaterials Pumped at Telecomm Wavelengths

Liang Luo, Ioannis Chatzakis, Jigang Wang[*], Fabian B. P. Niesler, Martin Wegener, Thomas Koschny, and Costas M. Soukoulis[*]

* To whom correspondence should be addressed. E-mail: C. M. S. (soukoulis@ameslab.gov); J. W. (jgwang@iastate.edu)

Optical Measurement:

In our experiment, a Ti:Sapphire amplifier is used with center wavelength 800 nm, pulse duration 35 fs at 1 kHz repetition rate. The main portion of the output from the amplifier is used to pump the optical parametric amplifier (OPA) to produce tunable near-infrared radiation from 1100-2600 nm with ~140 fs pulse duration, which serves as a generation beam to pump the SRR/ZnTe emitter. The generated THz pulses are then focused onto the ZnTe or GaSe detector by a parabolic mirror. Residual scattered NIR radiation from the pump pulse is removed from the THz signal by a Teflon filter before the detector. The other small portion of the output from the amplifier is used as a sampling beam to detect the THz pulses via the electro-optic sampling of the detector crystals. The THz section of the setup is purged with dry $N_2$ gas. Note the wire grid polarizer shown in Fig. 1a indicates the position where the polarization of generated THz pulses is measured. It is used only for the determination of the polarization and not presented for all the other experiments discussed in the paper. For polarization dependence measurement, the SRR emitter is rotated while the near-infrared pump beam remains fixed, so that the polarization of generated THz pulses is rotated together with SRR emitter accordingly. In addition, the detection ZnTe crystal is also polarized, which explains the $\cos^3(\theta)$ fitting used in Fig. 4.

Next, we describe the procedure to extract the second-order nonlinear susceptibility of a single layer SRR emitter. We start with the wave equation in a nonlinear medium propagating in z-axis as[1]:

$$\frac{\partial^2 E_T(z,t)}{\partial z^2} - \frac{n_T^2}{c^2}\frac{\partial^2 E_T(z,t)}{\partial t^2} = \frac{1}{\varepsilon_0 c^2}\frac{\partial^2 P_T^{(2)}(z,t)}{\partial t^2} = \frac{\chi^{(2)}}{c^2}\frac{\partial^2 |E_o(z,t)|^2}{\partial t^2} \tag{1}$$

where $E_T(z,t)$ is the generated THz field, $n_T$ the refractive index of THz pulses in the nonlinear medium, $\varepsilon_0$ the free space permittivity, $c$ the speed of light, $P_T^{(2)}(z,t)$ the second-order polarization of the nonlinear medium due to the optical pump beam $E_o(z,t)$, and $\chi^{(2)}$ the second-order nonlinear susceptibility of the nonlinear medium. As shown in supplementary Fig. S1, we define a Gaussian optical pump pulse propagating in a nonlinear medium, and at z=z the optical pulse field amplitude can be expressed as $E_o(z',t) = E_0 e^{-a(t-z'/v_0)^2}$, where $a = 2\ln 2/\tau^2$ and $\tau$ is the optical pulse duration, which is 140 fs in our case, and $v_o$ is the group velocity of the optical pump pulse in the medium. THz radiation generated from an infinitesimal thin layer of the nonlinear medium z=z can be expressed as

$$E_T(z',t) = A\chi^{(2)}(1-4ax^2)e^{-2ax^2},$$

where $x = t - \frac{z'}{v_o} - \frac{L-z'}{v_T}$, A is a constant, L the thickness of the medium, and $v_T$ the phase velocity of THz pulses in the medium. Consequently, the THz field generated from a single sheet of SRRs can be obtained as

$$E_T^{SRR}(t) = A\chi_{SRR}^{(2)}(1-4at^2)e^{-2at^2} \tag{2}$$

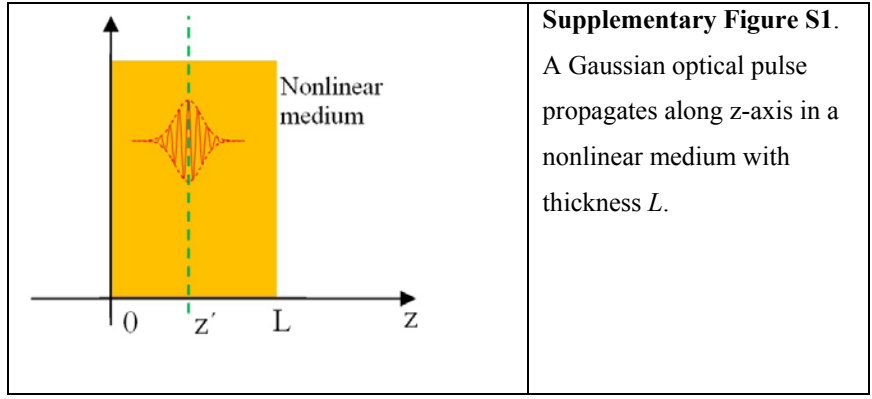

**Supplementary Figure S1**. A Gaussian optical pulse propagates along z-axis in a nonlinear medium with thickness $L$.

In a thick nonlinear medium, the THz field generated can be obtained by taking the integration of the sheet result over the thickness. For a 0.2 mm thick ZnTe crystal, we have:

$$\begin{aligned}
E_T^{\text{ZnTe}}(t) &= \int_0^{L=0.2\text{mm}} E_T(z',t)\mathrm{d}z' = \int_0^{L=0.2\text{mm}} A\chi^{(2)}_{\text{ZnTe}}(1-4ax^2)e^{-2ax^2}\mathrm{d}z' \\
&= \frac{Ac\chi^{(2)}_{\text{ZnTe}}}{n_T - n_o} \int_{t-2.10\text{ps}}^{t-1.89\text{ps}} (1-4ax^2)e^{-2ax^2}\mathrm{d}x \\
&= A\chi^{(2)}_{\text{ZnTe}}(9.1\times10^{-4}\text{m/ps})[(t-1.89\text{ps})e^{-2a(t-1.89\text{ps})^2} - (t-2.10\text{ps})e^{-2a(t-2.10\text{ps})^2}]
\end{aligned} \quad (3)$$

In the calculation, the group refractive index of optical pump beam at 1500 nm is[2] $n_o = n_{gr}(1500\text{ nm}) = 2.82$. The central frequency of the THz spectrum ~ 2 THz is used to determine the refractive index of THz in ZnTe, *i.e.*, $n_T = n_T(2\text{ THz})$. Because we found that our calculation is very sensitive to the indices which depend on crystal specifications, so in order to get as accurate calculation as possible, we measured it experimentally by comparing the time domain THz transmission through free space and the 0.2 mm ZnTe crystal, from where the refractive index is obtained. The results are shown in supplementary Fig. S2, which indicates that $n_T = 3.15$ at 2 THz.

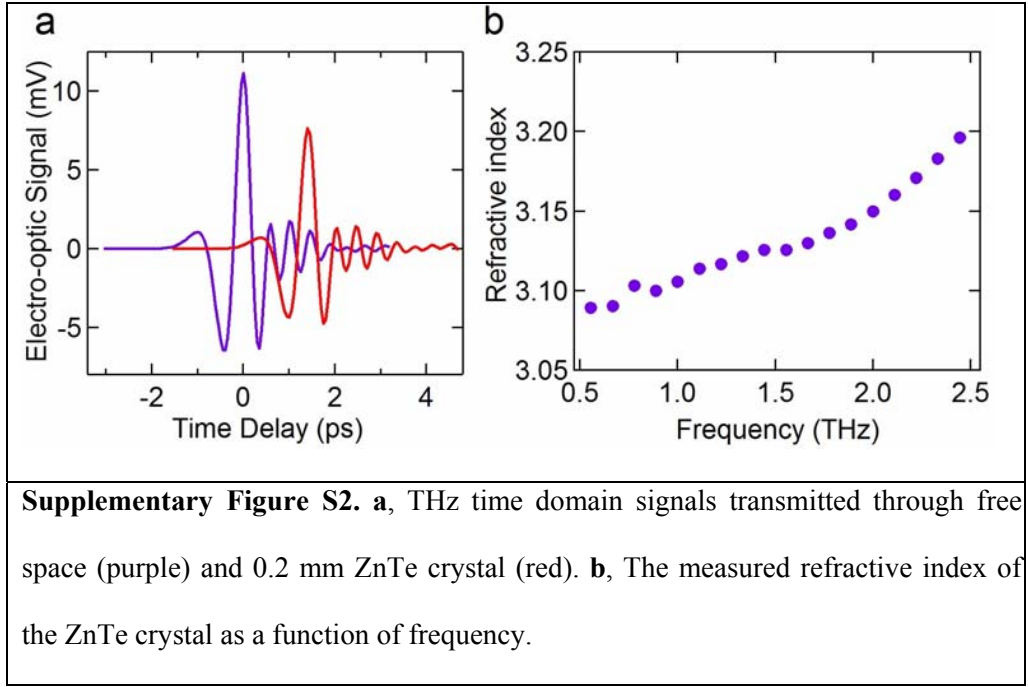

**Supplementary Figure S2. a**, THz time domain signals transmitted through free space (purple) and 0.2 mm ZnTe crystal (red). **b**, The measured refractive index of the ZnTe crystal as a function of frequency.

Knowing the measured peak-to-peak amplitudes $\Delta E_T^{\text{SRR}} \approx 0.2 \times \Delta E_T^{\text{ZnTe}}$ and $\chi_{\text{ZnTe}}^{(2)} = 2r_{41} = 8 \times 10^{-12}$ m/V, where $r_{41} = 4 \times 10^{-12}$ m/V [3,4] is the electro-optic coefficient of ZnTe, we can extract $\chi_{\text{SRR}}^{(2)}$ from calculated THz electric fields using equations (2) and (3) as shown in supplementary Fig. S3, where $\chi_{\text{SRR}}^{(2)} = 0.8 \times 10^{-16}$ m$^2$/V reproduces the experimental peak-to-peak ratio. Therefore, we reveal a gigantic resonant sheet nonlinear susceptibility of SRRs, which is three orders of magnitude higher than the typical surface and sheet values. For example, typical surface values are in the range of $10^{-22}$ m$^2$/V ~ $10^{-21}$ m$^2$/V for fused silica or BK7 glass, ~$10^{-19}$ m$^2$/V for liquid crystals, inorganic and organic thin films[5-7].

Note the above calculations do not take into account the detector response function since THz waves from both SRRs and ZnTe are measured by the same detector.

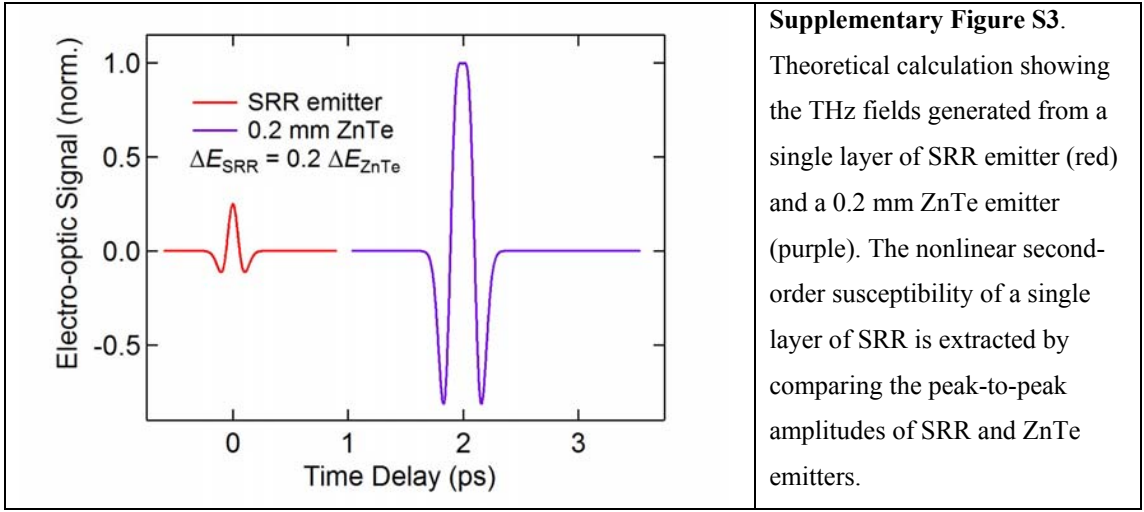

**Supplementary Figure S3.** Theoretical calculation showing the THz fields generated from a single layer of SRR emitter (red) and a 0.2 mm ZnTe emitter (purple). The nonlinear second-order susceptibility of a single layer of SRR is extracted by comparing the peak-to-peak amplitudes of SRR and ZnTe emitters.

Theory:

The observed THz emission from the SRR sample is due to optical rectification (OR) by the second order non-linear electric response arising from the electrons in the metal that makes up the SRR.

As shown in previous literature, the electron gas in the metal can be described by a hydrodynamic model[8-10] know as Maxwell-Vlassov theory, which goes beyond the usual linear Drude model for the metallic response and contains second order non-linearities:

$$\partial_t \mathbf{j} = \frac{ne^3}{m}\mathbf{E} - \gamma\mathbf{j} - \frac{e}{m}(\mathbf{E}\,\mathrm{div}\mathbf{E} + \mathbf{j}\times\mathbf{B}) + \frac{1}{ne}(\mathbf{j}\cdot\mathrm{grad})\mathbf{j} + \frac{e}{m}\mathrm{grad}\,p \;,$$

$$p(n) = \frac{1}{5}(3\pi^2)^{2/3}\frac{\hbar}{m}n^{5/3}$$

The first two terms comprise the linear response as described by the conventional Drude model, the following terms are the non-linear response of the electrons and represent Lorentz force, convective acceleration and Fermi pressure, respectively. All the nonlinear terms are second order in the Maxwell fields, giving rise to a non-zero second order electric polarizability $\chi^{(2)}$, which is responsible for 2nd harmonic generation as well as for the THz generation via OR considered here.

As was shown in Ref. 10, the main non-linear contributions are coming from the $\mathbf{E}\,\mathrm{div}\,\mathbf{E}$ and $(\mathbf{j}\cdot\mathrm{grad})\mathbf{j}$ terms, which do not contribute in the bulk but do contribute on surfaces. Qualitatively, both terms behave like $j\rho$, current times accumulated density on the surface: As a consequence the non-linear current is parallel or antiparallel to the linear current induced by the external excitation in regions of increasing or decreasing surface charge, respectively. This behaviour is illustrated in Fig. S4: In the case of a simple straight nanorod, shown in Fig. S4(a), the radiation caused by the non-linear current contribution (red) in the two regions of surface charge accumulation (i.e. both ends of the nanorod) is out of phase, interferes destructively, and is not observed in the far field. Similarly, the symmetry of the current distribution at the electric dipole resonance of the SRR shown in Fig. S4(b), i.e. with the external electric field along the two parallel arms of the SRR, leads to non-linear polarization that is parallel to the linear currents at the top ends of the SRR where (positive) charge is accumulating and anti-parallel at the bottom where (positive) charge is depleting. Hence, again, no radiation due to the non-linear current contributions is observed in the far field. If, however, we bend the nanorod into an U-shaped SRR and excite the resonant mode that has a continuous current (without nodes) around the SRR ring as shown in Fig. S4(c), i.e. from the tip of one arm to the tip of the other, the non-linear current in both arm are now parallel, their radiated fields interfere constructively and are observed in the far field. This mode is the magnetic resonance of the SRR, to which we couple electrically by the incident optical pulse (i.e. via the bi-anisotropic electric dipole moment across the gap), such that the resonance can be excited for normal incidence to the SRR plane. It should be noted that, although the mechanism also holds for second harmonic generation reported before[8,9], the current phase-resolved THz results allow us to solve the completely different pressing challenges in THz opto-electronics and gain new insights in nonlinear optics of metamaterials, i.e., quantitatively reveal the second order susceptibility of the SRRs and phase reversal of THz emission, which have not been obtained in prior the SHG experiment.

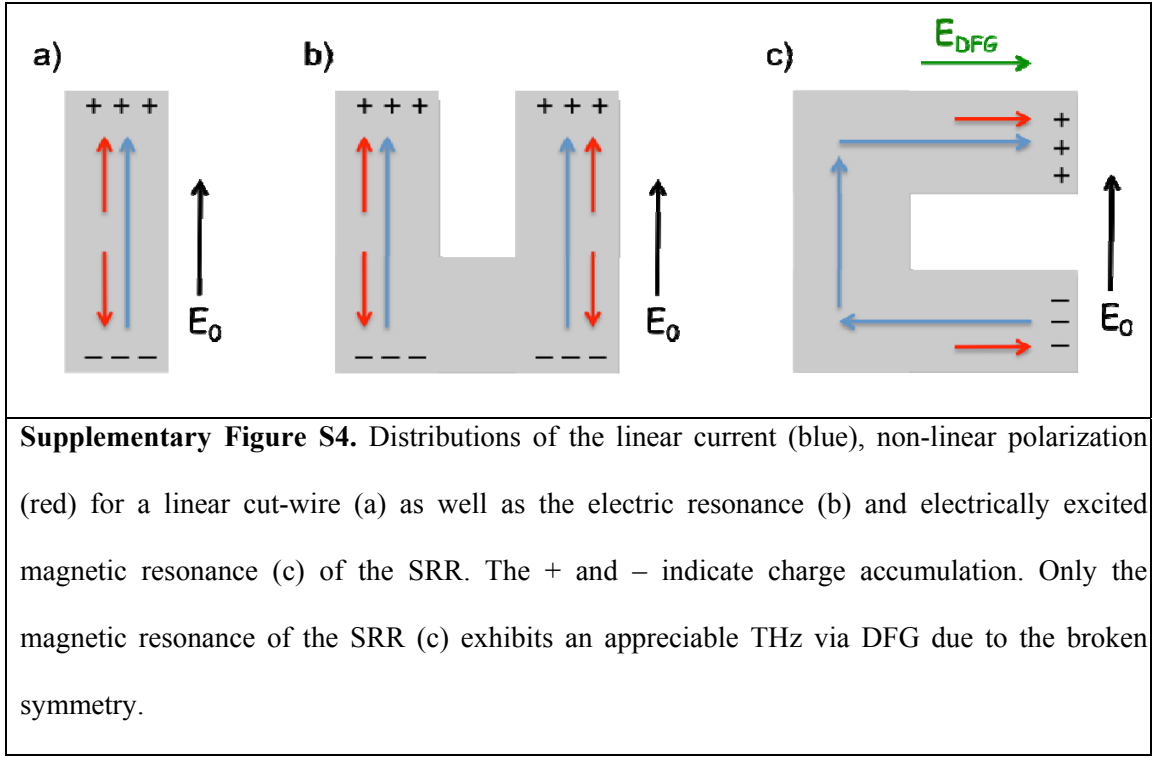

**Supplementary Figure S4.** Distributions of the linear current (blue), non-linear polarization (red) for a linear cut-wire (a) as well as the electric resonance (b) and electrically excited magnetic resonance (c) of the SRR. The + and − indicate charge accumulation. Only the magnetic resonance of the SRR (c) exhibits an appreciable THz via DFG due to the broken symmetry.

In first approximation we can describe the temporal optical input signal as a linearly chirped Gaussian pulse,

$$G(t) = e^{i(\psi t^2 + \Omega t + \phi)} h_{\sigma_0}(t) + e^{-i(\psi t^2 + \Omega t + \phi)} h^{\star}_{\sigma_0}(t), \quad \psi, \Omega, \phi, \sigma_o \in \mathbb{R}, \; \sigma_o > 0;$$

$$\text{where} \quad h_\sigma(t) = \tfrac{1}{2} e^{-\tfrac{1}{2}\sigma^2 t^2} \Leftrightarrow h_\sigma(\omega) = (2\sigma\sqrt{2\pi})^{-1} e^{-\tfrac{\omega^2}{2\sigma^2}} \quad \text{and} \quad \mathrm{Re}\,\sigma^2 > 0.$$

The linear chirp $\alpha = 2\psi/\sigma_0^2$ is just a redefinition $\sigma = \sigma_o \sqrt{1 - i\alpha}$ and can be obtained from the difference in spectral width of the measured power spectrum of the pump pulse ($\sigma \approx 2\pi \times 9$ THz) and the temporal width of the pump pulse derived from a cross-correlation measurement ($\sigma_0 \approx 2\pi \times 2$ THz). $\Omega$ is the center frequency of the pump pulse and $\phi$ a phase shift between carrier and envelope.

$$G(t) = e^{i(\Omega t + \phi)} h_\sigma(t) + e^{-i(\Omega t + \phi)} h^{\star}_\sigma(t) \Leftrightarrow$$

$$G(\omega) = e^{i\phi} h_\sigma(\omega - \Omega) + e^{-i\phi} h^{\star}_\sigma(-\omega - \Omega) \quad \text{where} \quad \sigma = \sigma_o\sqrt{1 - i\alpha}, \; \alpha = \frac{2\psi}{\sigma_o^2}.$$

with the second order non-linear response being proportional to

$$G^{(2)}(t) = 2h_{\sigma_o}^2(t) + e^{i(2\Omega t+2\phi)}h_\sigma^2(t) + e^{-i(2\Omega t+2\phi)}\left[h_\sigma^2(t)\right]^\star \Leftrightarrow$$

$$G^{(2)}(\omega) = \frac{1}{2}\left(2h_{\sqrt{2}\sigma_o}(\omega) + e^{2i\phi}h_{\sqrt{2}\sigma}(\omega-2\Omega) + e^{-2i\phi}h_{\sqrt{2}\sigma}^\star(-\omega-2\Omega)\right).$$

The OR is the first term in the spectrum above; the other two summands represent the SHG, which is removed from the signal by the Teflon filter in the THz signal path and by the detector crystals acting as an effective low-pass filter. Note that the chirp drops out and the OR signal only depends on the temporal envelope of the pump pulse.

We can write the radiated THz field in terms of a $\chi^{(2)}$ polarizability of the SRR sample:

$$\mathbf{E}_{rad}^{(nl)}(\omega) \sim \chi^{(2)}(-i\omega)^2\, G^{(2)}(\omega) \Leftrightarrow$$

$$\mathbf{E}_{rad}^{(nl)}(t) \sim \chi^{(2)}\, \mathcal{F}^{-1}\left[-(-i\omega)^2\, G^{(2)}(\omega)\right](t) \sim -\chi^{(2)}\, \partial_t^2 h_{\sqrt{2}\sigma_0}(t).$$

The time derivative in the radiated fields suppresses zero-frequency components such that the THz spectrum has a peak at finite frequency,

$$\mathbf{E}_{rad}^{(nl)}(\omega) \sim \chi^{(2)}\omega^2 e^{-\frac{\omega^2}{4\sigma_0^2}} \Leftrightarrow \mathbf{E}_{rad}^{(nl)}(t) \sim \chi^{(2)}\sigma_0^2(1-2\sigma_0^2 t^2)e^{-\sigma_0^2 t^2}.$$

The THz spectrum peaks at $\omega = 2\sigma_0$ and has a bandwidth $\Delta\omega_{\text{FWHM}} \approx 2.31\sigma_0$.

Strictly, the radiated non-linear fields are given by the currents in the SRR, which are proportional to the reflection amplitude of the electric sheet given by the SRR metasurface. So in the formulas above we have $G(\omega) \sim R(\omega)G^{(in)}(\omega)$. From the linear transmittance measurements of the SRR we get $R(\omega) \approx Z\sigma_e(\omega)/[2+Z\sigma_e(\omega)]$ with $Z\sigma_e(\omega) \approx -i\alpha\omega/(\beta^2-\omega^2-i\gamma\omega)$ where $\alpha \approx 384$, $\beta \approx 2\pi\times 199$ THz, and $\gamma \approx 2\pi\times 15.4$ THz. Thus, since the SRR resonance is very wide compared to the bandwidth of the optical pump pulse envelope ($\sigma_0 \approx 2\pi\times 2$ THz) the effect of $R(\omega)$ on the THz pulse shape is negligible: In our experiments, the achievable THz bandwidth is limited by the duration of the optical pump pulse, not the SRR response.

A second, more severe problem is that the bandwidth and, in particular, the upper cut-off frequency (low-pass, between 2 and 3 THz depending on the used crystal) of the detectors is much smaller than the expected THz signal. As a consequence, the spectrum of the observed THz signal in our experiments is not limited by the OR but essentially given by the bandwidth of the detectors.

The so predicted THz waveform agrees well with the experimental observation.